\documentclass[twoside]{ae100prg}
\usepackage{graphicx}
\usepackage[breaklinks]{hyperref}
\usepackage{booktabs}

\begin{document}
\title{A reference for the covariant Hamiltonian boundary term}


\author{James M. Nester$^{1,2}$, Chiang-Mei Chen$^{1}$
 Jian-Liang Liu$^1$, and Gang Sun$^1$}

\address{$^1$ Department of Physics \& Center for Mathematics and Theoretical Physics, National Central University, Chungli 320 Taiwan}
\address{$^2$ Graduate Institute of Astronomy, National Central University, Chungli 320 Taiwan}

\email{nester@phy.ncu.edu.tw, cmchen@phy.ncu.edu.tw, liujl@phy.ncu.edu.tw, liquideal@gmail.com}

\begin{abstract}
The Hamiltonian for dynamic geometry generates the evolution
of a spatial region along a vector field. It includes a boundary term
which determines both the value of the Hamiltonian and the boundary
conditions. The value gives the quasi-local quantities:
energy-momentum, angular-momentum/center-of-mass. The boundary term
depends not only on the dynamical variables but also on their
reference values; the latter determine the ground state (having
vanishing quasi-local quantities). For our preferred boundary term for
Einstein's GR we propose 4D isometric matching and extremizing the
energy to determine the reference metric and connection values.
\end{abstract}

\section{Introduction}

Energy-momentum is the source of gravity.
Gravitating bodies can exchange energy-momentum with gravity---{\em locally}---yet there is no well defined energy-momentum density for gravity itself.  This inescapable conclusion 
can be understood as a consequence of the equivalence principle (for a discussion see~\cite{MTW73}, Section 20.4).

\section{Quasi-local energy-momentum}

The standard approaches aimed at identifying an energy-momentum density for gravitating systems always led to various non-covariant, reference frame dependent, energy-momentum complexes (such expressions are generally referred to as pseudotensors).  There are two types of ambiguity. First, there was no unique expression, but rather many that were found by various investigators---including
Einstein~\cite{Traut62}, Papapetrou~\cite{Papa48}, Landau-Lifshitz~\cite{LL62}, Bergmann-Thompson~\cite{BT53}, M{\o}ller~\cite{Mol58}, Goldberg~\cite{Gol58}, and Weinberg~\cite{Wein72}---so which expression should  be used?
And second---in view of the fact all of these expressions are inherently reference frame dependent---for a chosen expression which reference frame should be used to give the proper physical energy-momentum localization?

The more modern idea is {\em quasi-local}, i.e., energy-momentum should be associated not with a local density but rather with a closed 2-surface;  for a comprehensive review see~\cite{Sza09}.

One approach to energy-momentum is via the {\em Hamiltonian} (the generator of time evolution).
It turns out that this actually includes all the classical pseudotensors as special cases, while taming their ambiguities---providing clear physical/geometric meaning~\cite{CNC99,Nester04}.

\section{The covariant Hamiltonian formulation results}

We have developed a covariant Hamiltonian formalism that is applicable to a large class of geometric gravity theories~\cite{CNT95,CN99,CNC99,CN00,Nester04,CNT05,Nester08}.
For such theories the Hamiltonian 3-form ${\cal H}(N)$ is both a conserved Noether current,
\begin{equation}d{\cal H}(N)\propto \hbox{field eqns}\simeq 0\,,\end{equation}
 as well
as the generator of the evolution of a spatial region along a space-time displacement vector field. It has the general form
\begin{equation}{\cal H}(N)
=  N^\mu {\cal H}_\mu + d {\cal B}(N)\,,   \end{equation}
where $N^\mu{\cal H}_\mu$---which generates the evolution equations---is itself proportional to certain field equations (initial value constraints) and thus vanishes ``on shell''.  Consequently
 the {\em value} of the Hamiltonian is determined by the total differential (boundary) term:
\begin{equation}
E(N,\Sigma):=\int_\Sigma{\cal H}(N)=\oint_{\partial\Sigma}{\cal B}(N)\,.
\end{equation}
Thus, the value is {\em quasi-local}.  From this boundary term, with suitable choices of the vector field on the boundary, one can determine the quasi-local energy-momentum and angular momentum/center-of-mass.

 It should be noted that the boundary 2-form ${\cal B}(N)$ can be modified---{\em by hand}---in any way without destroying the conservation property.  (This is a particular case of the usual Noether conserved current ambiguity.)  With this freedom one can arrange for almost any conserved quasi-local values.
Fortunately the Hamiltonian's role in generating evolution equations tames that freedom.

\section{Boundary variation principle, reference values}

One must look to the boundary term in the {\em variation of the Hamiltonian} (see~\cite{Lan49,RT74,KT79}).
Requiring it to vanish yields the {\em boundary conditions}.
The Hamiltonian is {\em functionally differentiable} on the phase space
of fields {\em satisfying these boundary conditions}.
Modifying the boundary term changes the boundary conditions.
(The different classical pseudotensors are each associated with a specific ``superpotential'' which can serve as the Hamiltonian boundary term,  thus they correspond to Hamiltonians with different boundary conditions~\cite{CNC99}.)

In order to accommodate suitable boundary conditions
        one must, in general, also introduce certain
       {\em reference values} which represent the
       ground state of the field---the ``vacuum'' (or background field) values.
To this end for any quantity $\alpha$ we let $\Delta\alpha:=\alpha-\bar\alpha$ where $\bar\alpha$ is the reference value.

\section{Preferred boundary term for GR}

Some time ago we identified for GR two {\em covariant-symplectic} boundary terms~\cite{CNT95}; one, which was
also found\footnote{Via a different route, using a Noether type argument with a global reference.}
 at about the same time by  Katz, Bi{\v c}\'ak and Lynden-Bell~\cite{LBKB95,KBLB97},
has become our preferred choice:\footnote{Here $\Gamma^\alpha{}_\beta$ is the connection one-form,
$\eta^{\alpha\beta\dots} := * (\vartheta^\alpha \wedge \vartheta^\beta\wedge \cdots)$ and $i_N$
denotes the interior product (aka contraction) with the vector field $N$.}
\begin{eqnarray}{\cal B}(N)=\frac{1}{2\kappa}(\Delta\Gamma^{\alpha}{}_{\beta}\wedge
i_N\eta_{\alpha}{}^{\beta}+\bar
D_{\beta}N^\alpha\Delta\eta_{\alpha}{}^\beta)\,, \label{B}
 \end{eqnarray}
This choice corresponds to fixing  the orthonormal coframe $\vartheta^\mu$ (equivalently the metric) on the boundary:
\begin{equation}
\delta{\cal H}(N)\sim di_N(\Delta\Gamma^\alpha{}_\beta\wedge\delta\eta_{\alpha}{}^\beta)\,.
\end{equation}
Like other choices, at spatial infinity it gives the ADM~\cite{ADM60}, MTW~\cite{MTW73}, Regge-Teitelboim~\cite{RT74}, Beig-\'O Murchadha~\cite{BoM87}, Szabados~\cite{Sza03,Sza06} energy, momentum, angular-momentum, center-of-mass.

Its special virtues include
{(i)} at {null infinity} it directly gives
  the Bondi-Trautman energy and the Bondi energy flux~\cite{CNT05},
{(ii)} it is ``covariant'',
{(iii)} it has a positive energy property,
{(iv)} for small spheres it gives a positive multiple of the Bel-Robinson tensor,
{(v)} it yields the first law of thermodynamics for black holes~\cite{CN99},
{(vi)}  for spherically symmetric solutions it has the {hoop} property~\cite{MTXM10}.

\section{The reference and the quasi-local quantities}

 For all other fields it is {appropriate} to choose  {vanishing reference values} as the reference ground state---the vacuum.
But for geometric gravity the standard ground state is the {non-vanishing} Minkowski metric.
Thus a non-trivial reference is {essential}.

 Using standard Minkowski coordinates $y^i$, a Killing field of the reference has the form
$N^k=N^k_0+\lambda_0^k{}_l y^l$, where the translation parameters $N_0^k$ and the boost-rotation parameters $\lambda_0^{kl}=\lambda_0^{[kl]}$ are constants.
 The 2-surface integral  of the Hamiltonian boundary term then gives a value of the form
\begin{equation}\oint_S{\mathcal B}(N)=-N_0^kp_k(S)+\frac12 \lambda_0^{kl}J_{kl}(S)\,,\end{equation}
 which yields not only a quasi-local {energy-momentum} but also a quasi-local {angular momentum/center-of-mass}.  The integrals $p_k(S),\ J_{kl}(S)$ in the spatial asymptotic limit agree with accepted expressions for these quantities~\cite{MTW73,RT74,BoM87,Sza03,Sza06}.

\section{The reference}

  For energy-momentum one takes $N$ to be a translational Killing field of the Minkowski reference.  Then the second term in our quasi-local boundary expression~(\ref{B}) vanishes.
Let us note in passing that holonomically (with vanishing reference connection coefficients) the first term in~(\ref{B}) reduces to Freud's 1939 superpotential~\cite{Freud39}.
Thus we are in effect here making a proposal for good coordinates for the Einstein pseudotensor.

To construct a reference,
 choose, in a neighborhood of the desired spacelike boundary 2-surface $S$, four smooth functions $y^i,\ i=0,1,2,3$  with $dy^0\wedge dy^1\wedge dy^2\wedge dy^3\ne0$; they define a Minkowski reference by $\bar g=-(dy^0)^2+(dy^1)^2+(dy^2)^2+(dy^3)^2$.
This is equivalent to finding a diffeomorphism for a neighborhood of the 2-surface into Minkowski space. The reference connection can now be obtained from the pullback of the flat Minkowski connection.

With constant $N^k$ our quasi-local expression now takes the form
\begin{equation}{\mathcal B}(N)=N^k x^\mu{}_k(\Gamma^\alpha{}_\beta-x^\alpha{}_j\,dy^j{}_\beta)\wedge\eta_{\mu\alpha}{}^\beta\,,\label{B2}
\end{equation}
where $dy^k=y^k{}_\alpha dx^\alpha$ has the inverse $dx^\alpha=x^\alpha{}_k dy^k$.

\section{Isometric matching of the 2-surface}

The reference metric on the dynamical space has the components
\begin{equation}
\bar g_{\mu\nu}=\bar g_{ij}y^i{}_\mu y^j{}_\nu\,.
\end{equation}
Consider the usual embedding restriction: isometric matching of the 2-surface $S$.
This can be expressed quite simply in terms of quasi-spherical foliation adapted coordinates $t,r,\theta,\varphi$ as
\begin{equation}
g_{AB}\dot=\bar g_{AB}=\bar g_{ij}y^i_A y^j_B=-y^0_A y^0_B+\delta_{ij}y^i_A y^j_B\,,
\end{equation}
where $S$ is given by constant values of $t,r$, and $A,B$ range over $\theta,\varphi$.  We use $\dot=$ to indicate a relation which holds only on the 2-surface $S$.

From a classic closed 2-surface into $\mathbb R^3$ embedding theorem---as long as one restricts $S$ and $y^0(x^\mu)$ so that on $S$
\begin{equation}
g_{AB}':=g_{AB}+y^0_A y^0_B
\end{equation}
is convex---one has a unique embedding.
Wang and Yau have discussed in detail this type of embedding of a 2-surface into Minkowski controlled by one function in their recent quasi-local work~\cite{WYcmp09,WYprl09}.

\section{Complete 4D isometric matching}

Our ``new'' proposal is:
complete 4D isometric matching on $S$.
 (We remark that this was already suggested by Szabados back in 2000\footnote{At a workshop in Hsinchu, Taiwan.}, and he has since extensively explored this idea~\cite{Sza05} in unpublished work.)

Complete 4D isometric matching imposes 10 constraints,
$$ g_{\mu\nu}|_S\dot=\bar g_{\mu\nu}|_S\dot=\bar g_{ij}y^i{}_\mu y^j{}_\nu|_S\,,
$$
on the 16 $y^i_\alpha(t_0,r_0,\theta,\varphi)$.  On the 2-surface $S$ these 16 quantities are actually determined by 12 independent embedding functions: $y^i,y^i{}_t,y^i{}_r$ (since from $y^i$ on $S$ one can get $y^i{}_\theta,y^i{}_\varphi$).  There remain $2=12-10$ degrees of freedom in choosing the reference.

One could as an alternative use orthonormal frames. Then the 4D isometric matching can be represented by $\vartheta^\alpha\dot=\bar\vartheta^\alpha$.  But the reference coframe has the form $\bar\vartheta^\alpha=dy^\alpha$. Thus one should Lorentz transform the coframe $\vartheta^\alpha$ to match $dy^\alpha$ on the 2-surface $S$.
This leads to an integrability condition: the 2-forms $d\vartheta^\alpha$ should vanish when restricted to the 2-surface:
\begin{equation}d\vartheta^\alpha|_S\dot=0,\end{equation}
This is 4 conditions restricting the 6 parameter local Lorentz gauge freedom.  Which again reveals that after 4D isometric matching there remains $2=6-4$ degrees of freedom in choosing our reference.
By the way, this orthonormal frame formulation shows that our procedure can alternatively be viewed as finding a good frame for the ``teleparallel gauge current''~\cite{dAGP00}.

\section{The best matched reference geometry}

There are
 12 embedding variables subject to 10 4D isometric matching conditions, or
equivalently, 6 local Lorentz gauge parameters subject to 4 frame embedding conditions.
To fix the remaining 2, one can regard the quasi-local value as a measure of the difference between the dynamical and the reference boundary values.
So we propose taking the optimal ``best matched'' embedding as the one which gives the extreme value to the associated invariant mass $m^2=-p_i p_j \bar g^{ij}$.
This is reasonable, as one expects the  quasi-local energy to be non-negative and to vanish only for Minkowski space.

More precisely, we note two different situations:\\
{\bf I:} Given a 2-surface $S$ find the critical points of $m^2$.
This should determine the reference up to Poincar\'e transformations.\\
{\bf II:} Given a 2-surface $S$ and a vector field $N$, then one can look to the choices of the embedding variables that are a critical point of $E(N,S)$.  (Afterward one could extremize over the choice of $N$.)

Based on some physical and practical computational arguments it seems reasonable to expect a unique solution in general.

For spherically symmetric systems (both static and dynamic), using this and some other related strategies we have found reasonable quasi-local energy results~\cite{WCLN10,LCN11,WCLN11,WCLN12}.


\section*{Acknowledgements} We much benefited from discussing some of this material with M.-F. Wu. This work was supported by the National Science Council of the R.O.C. under the
grants NSC-100-2119-M-008-018, NSC-101-2112-M-008-006 (JMN) and NSC 99-2112-M-008-005-MY3
(CMC) and in part by the National Center of Theoretical Sciences (NCTS).

\section*{References}
\bibliography{ae100prg121013jmn}

\end{document}